\documentclass[letterpaper, 10 pt, journal, twoside]{IEEEtran}  

\IEEEoverridecommandlockouts                              

\title{Comments on Truncation Errors\\for Polynomial Chaos Expansions}
\author{Tillmann M\"uhlpfordt,$^{a,\star}$ Rolf Findeisen,$^{b,\star\star}$ Veit Hagenmeyer,$^{a,\star}$ Timm Faulwasser$^{a,\star}$
	\thanks{$^\star$TM, VH, and TF acknowledge support  by the  Helmholtz Association under the Joint Initiative ``Energy System 2050 -- A Contribution of the Research Field Energy''.
		TF acknowledges support from the Baden-W\"urttemberg Stiftung under the Elite Programme for Postdocs.}
	\thanks{$^{\star\star}$RF acknowledges support by BMBF project ``InTraSig,'' 031A300A.}
\thanks{$^{a}$Institute for Applied Computer Science, Karlsruhe Institute of Technology, Germany, $\{$tillmann.muehlpfordt, \,veit.hagenmeyer, timm.faulwasser$\}$@kit.edu.  
	}
\thanks{$^b$Laboratory for Systems Theory and Automatic Control, Otto-von-Guericke University Magdeburg, Germany, rolf.findeisen@ovgu.de}%
}

\usepackage{amssymb}
\usepackage{amsmath}
\usepackage[english]{babel}
\usepackage{xspace}
\usepackage{cases}
\usepackage{color}
\usepackage{graphicx}
\usepackage{epstopdf}
\usepackage{upgreek}
\usepackage[caption=false]{subfig}

\usepackage{enumitem}
\usepackage{booktabs}
\usepackage{array}
\usepackage{blindtext}

\usepackage[hidelinks]{hyperref}

\newcommand{\QED}{\hfill {$\blacksquare$}}
\newcommand{\argmin}[1]{\underset{#1}{\operatorname{argmin~}}}

\newtheorem{problem}{Problem}
\newtheorem{ass}{Assumption}
\newtheorem{defi}{Definition}
\newtheorem{thm}{Theorem}

\newtheorem{coro}{Corollary}

\newtheorem{lemm}{Lemma}

\newtheorem{rema}{Remark}

\newtheorem{exmpl}{Example}

\newcommand{\movefig}{-7mm}

\newcommand{\nxi}{n_\xi}

\renewcommand{\arraystretch}{1.15}
\newcommand{\pdf}{\textsc{pdf}\xspace}
\newcommand{\pdfs}{\textsc{pdf}s\xspace}

\newcommand{\Ltwospace}{\mathrm{L}^2(\Omega,\mu;\mathbb{R})}

\newcommand{\arguargu}{z}
\newcommand{\argu}{\rv{\arguargu}}
\newcommand{\Argu}{Z}

\newcommand{\pce}{\textsc{pce}\xspace}

\newcommand{\mpc}{\textsc{mpc}\xspace}
\newcommand{\qp}{\textsc{qp}\xspace}
\newcommand{\qps}{\textsc{qp}s\xspace}
\newcommand{\lti}{\textsc{lti}\xspace}
\newcommand{\lqr}{\textsc{lqr}\xspace}

\newcommand{\bs}[1]{\boldsymbol{#1}}
\newcommand{\rv}[1]{\mathsf{#1}}

\newcommand{\her}[1]{\mathrm{He}_{#1}}
\newcommand{\eaug}[2]{\tilde{e}_{#2}^{#1}}
\newcommand{\proj}[2]{\mathrm{P}_{\!#2}#1}

\newcommand{\kkt}{\textsc{kkt}\xspace}
\newcommand{\Nx}{n_{\chi}}
\newcommand{\Ncon}{n_{\text{con}}}
\newcommand{\Nact}{n_{\text{act}}}
\newcommand{\licq}{\textsc{licq}\xspace}

\newcommand{\Mact}{M_{\mathcal{A}}}

\newlist{mylist}{enumerate}{1}
\setlist[mylist]{label=\textsc{q}\oldstylenums{\arabic*}}
\newlist{mylistlist}{enumerate}{1}
\setlist[mylistlist]{label=\textsc{a}\oldstylenums{\arabic*}}

\definecolor{till}{rgb}{1.0,0.0,0.0}
\definecolor{timm}{rgb}{.0,.0,.0} 
\definecolor{veit}{rgb}{.8,.1,.1}

\def\timm{\textcolor{timm}}

\usepackage[citestyle=numeric-comp, uniquename=false, sorting=none, maxbibnames=6, maxcitenames=2, doi=false, url=false, isbn=false, eprint=false, backend=bibtex]{biblatex}
\bibliography{Approxerrors_lit.bib}

\begin{document}

\maketitle	
\thispagestyle{empty}
\pagestyle{empty}

\begin{abstract}
Methods based on polynomial chaos expansion allow to approximate the behavior of systems with uncertain parameters by deterministic dynamics.
These methods are used in a wide range of applications, spanning from simulation of uncertain systems to estimation and control.
For practical purposes the exploited spectral series expansion is typically truncated to allow for efficient computation, which leads to approximation errors.
Despite the Hilbert space nature of polynomial chaos, there are only a few results in the literature that explicitly discuss and quantify these approximation errors.
This work derives error bounds for polynomial chaos approximations of polynomial and non-polynomial mappings. Sufficient conditions are established, which allow investigating the question whether zero  truncation errors can be achieved and which series order is required to achieve this. Furthermore, convex quadratic programs, whose argmin operator is a special case of a piecewise polynomial mapping, are studied due to their relevance in predictive control. Several simulation examples illustrate our findings.
\end{abstract}
\section{Introduction}
\IEEEPARstart{U}{ncertainty} is inherent to many applications. Considering and counteracting disturbances is becoming ever more important as systems are pushed to the boundaries of operation, for economic reasons or for increased interoperability. 
 By now, many strategies have been developed to predict and counteract disturbances and uncertainties.
 With respect to systems and control they span robust control \cite{Zhou1996}, stochastic, and robust model predictive control (\mpc) approaches \cite{Mesbah16,Rawlings2009MPC,limon2009input}. 

Various methods for uncertainty description, prediction and decision making under uncertainties exist \cite{Xiu10book,jaulin2001applied}. Besides stochastic uncertainty descriptions, deterministic uncertainty descriptions and bounds are often used.
In the deterministic setting,  uncertainties are typically described by bounded sets, leading to worst-case assumptions and worst-case predictions of the future system behavior.

Instead, stochastic approaches treat the uncertainty as a realization of a random variable (often continuous, second-order) that has to be propagated through given mappings, e.g. system dynamics, to obtain insights on the influence on the variable of interest, or  for control.
Recently, polynomial chaos expansion (\pce) has gained popularity in the field of systems and control to propagate stochastic uncertainty descriptions and to quantify their influence \cite{Mesbah14,Mesbah16,Paulson17a}.
\pce originates in the works of Norbert Wiener \cite{Wiener38}. In \pce the stochastic variables are replaced by an (infinite) sum of weighted orthogonal polynomials \cite{Xiu10book}. Using Galerkin projection, the approximated system is deterministic but of larger dimension than the original system.
This expanded system has been used, for example, to design linear controllers \cite{fisher09aut,Du17a,Shen17a}, and has been exploited in model predictive control \cite{Fagiano12a,Mesbah14,paulson2014,Lucia15,Muehlpfordt16a}. 

For the sake of computational tractability, it is necessary to truncate the infinite polynomial chaos expansion to finite order. While the use of \pce in the field of systems and control is steadily increasing, it is commonly and frequently assumed that:
 (i) the input uncertainty $\argu$ can be exactly described using finitely many \pce coefficients; (ii) the nonlinear function---denoted in the following by $f(\cdot)$---that maps $\argu$ to the desired output $\rv{y} = f(\argu)$ is known analytically;\footnote{Here, $f(\cdot)$ represents a generic mapping, e.g. the state transition map of a system of ordinary differential equations or of an \lti system in discrete or continuous time, a system of nonlinear algebraic equations, or the argmin-operator of a suitable convex optimization problem.} and
 (iii) the output $\rv{y}$ can be exactly realized by a finite number of \pce coefficients.
Moreover, whereas under these conditions \pce is exact (in the L$^2$-sense) in the limit, its truncation is often a trade-off between approximation accuracy and computational tractability. For stability and performance guarantees, however, bounding the approximation error is important. As such, stability and performance guarantees  derived for the approximated system do not necessarily apply to the original stochastic system.

To the best of the authors' knowledge, there is only a limited number of  results that consider \pce truncation errors  directly: In \cite{FieldJr04} illustrative examples evaluate the accuracy of \pce. Yet, the errors are not computed rigorously. Rather, they are studied via extensive simulations.
	Similarly, \cite{Debusschere04} list several numerical challenges when using \pce, including the potential need for large \pce dimensions.

For \mpc-specific applications of \pce, error bounds on the first- and second-order moments, which are polynomial functions of \pce coefficients, are established in \cite{Lucia17}. These results provide a deep insight, however, no bounds on the error of the underlying projections in Hilbert spaces are given.
The authors of \cite{Augustin08} provide an upper bound on the truncation error using a univariate Hermitian basis based on differentiability assumptions of $f(\cdot)$. Yet, these results do not easily carry over to other bases.

The main contribution of the present paper is to leverage well-established Hilbert space theory \cite{Luenberger69book} to the end of  quantifying truncation errors for $\rv{y}$ in the $\mathrm{L}^2$-sense for multivariate uncertainties considering applications in the field of systems and control.
We provide exact error descriptions instead of error bounds.
Considering polynomial and non-polynomial mappings $f(\cdot)$, we tackle the question of how to choose the output \pce dimension such that the truncation error vanishes. Moreover, we establish bounds that allow the computation of minimum \pce dimensions such that a user-specified error tolerance is met. Furthermore, we study truncation errors for convex quadratic programs due to their relevance  in (stochastic) model predictive control.
Illustrative examples accompany the findings.

The remainder is organized as follows:  Section \ref{sec:ProblemFormulation} introduces \pce and the tackled research questions. Section \ref{sec:PolTruncationErrors} establishes results on \pce truncation errors for polynomial mappings. Also, truncation errors for convex quadratic programs are derived as they typically appear in model predictive control. 
Section \ref{sec:NonPolTruncationErrors} derives error bounds for the non-polynomial case.

\section{Problem Formulation}\label{sec:ProblemFormulation}
We consider random variables $\rv{y}$ that are the image of random variables $\argu$ under the square-integrable mapping
\begin{equation}\label{eq:mapping}
\rv{y} = f(\argu).
\end{equation}
We assume that $\argu$ and $\rv{y}$  are real-valued second-order random variables with multivariate $\mathbb{R}^{\nxi}$-valued stochastic germ $\rv{\xi}$, cf. \cite{Sullivan15book, Xiu10book}.
The mapping  $f: \argu\mapsto\rv{y}$ can, for example, describe the state transition map for continuous-time or discrete-time systems subject to uncertainties in the system matrix.
Also, it can describe the influence of a set of uncertain parameters/initial conditions on the output. Describing the dependence of the random variable $\rv{y}$ on the properties of the random variable $\argu$ is in general challenging. One way to do this is via \pce, which allows any random variable with finite second-order moment to be represented as a (possibly infinite) series of weighted orthogonal polynomials \cite{Xiu10book,Mesbah14,Sullivan15book}. The main question of interest in the present paper is to quantify the approximation error made if the series expansion is terminated early.

\subsection{Polynomial Chaos Expansion}
We first focus on describing real-valued random variables~$\argu$  from the Hilbert space $\Ltwospace$ of equivalence classes of univariate real-valued second-order random variables given the probability space $(\Omega,\mathcal{F},\mu)$.\footnote{With slight abuse of terminology, we will denote the set of equivalence classes of random variables simply as the \emph{set of random variables}.} 
	First note that the Hilbert space over the product probability space 
	is (under mild technical assumptions) equivalent to the Hilbert space tensor product and given by
	\begin{equation}
	\Ltwospace := 
	\bigotimes_{i=1}^{\nxi} \mathrm{L}^2(\Omega_i, \mu_i; \mathbb{R}),
	\end{equation}
	where $\Omega = \Omega_1{\times} \cdots {\times} \Omega_{\nxi}$, and $\mu = \mu_1 {\otimes} \cdots {\otimes} \mu_{n_\xi}$ denotes the tensor product, cf. \cite{Sullivan15book}.\footnote{In the following $\xi$ stands for the stochastic germ spanning $\Ltwospace$.}
Assume that the set of $\nxi$-variate polynomials $\{ \phi_j \}_{j = 0}^{\infty}$ spans the space $\Ltwospace$ and satisfies the orthogonality relation for all $i,j \in \mathbb{N}_0$
\begin{equation}
\langle \phi_i, \phi_j \rangle := \int_{\Omega} \phi_i(\tau) \phi_j(\tau) \mathrm{d}\mu(\tau) = \delta_{ij} \|\phi_i \|^2, 
\end{equation}
with Kronecker-delta $\delta_{ij}$, and the induced norm $\| \cdot \| = \sqrt{ \langle \cdot, \cdot \rangle }$. This allows the \pce of $\argu$ to be defined as:

\begin{defi}[Polynomial chaos expansion]
The polynomial chaos expansion of a real-valued random variable $\argu \in \Ltwospace$ is
\begin{equation}
\label{eq:FourierFormula}
\argu = \sum_{j = 0}^{\infty} \arguargu_j \phi_j, \quad \text{with} \quad \arguargu_j = \frac{\langle \argu, \phi_j \rangle}{\langle \phi_j, \phi_j \rangle},
\end{equation}
where $\arguargu_j \in \mathbb{R}$ is called the $j$th \pce coefficient \cite{Sullivan15book, Xiu10book}.\hfill $\square$
\end{defi}

In practice, the truncated \pce of $\argu$ is often considered to allow for more efficient calculations:
\begin{defi}[Truncated polynomial chaos expansion]
The truncated \pce of $\argu$ is
\begin{equation}
\label{eq:TruncatedPCE}
\proj{\argu}{\ell} := \sum_{j=0}^{\ell} \arguargu_j \phi_j,
\end{equation}
where $\ell + 1$ is the dimension of the subspace $\mathrm{\Argu} \subseteq \Ltwospace$ spanned by $\{ \phi_j \}_{j=0}^{\ell}$; \pce dimension in short.
The basis is chosen to contain all $\nxi$-variate polynomials $\phi_j$ of degree at most $d$ (in lexicographical order), yielding
\begin{equation}
\label{eq:PCEDimensionFormula}
\ell + 1 = \frac{(\nxi+d)!}{\nxi ! d!}.
\end{equation}
If the \pce coefficients are computed using the Fourier quotient \eqref{eq:FourierFormula}, the truncated \pce  $\proj{\argu}{\ell}$ from \eqref{eq:TruncatedPCE} is the orthogonal projection of $\argu$ onto $\mathrm{\Argu}$. \hfill $\square$
\end{defi}
Often, the stochastic germ $\xi$ is chosen to follow a Gaussian, Beta, Gamma, or uniform distribution (or a tensorized combination thereof) \cite{Mesbah14,Fagiano12a,Kim13a,Muehlpfordt17a,Mesbah16,Lucia15}.
Note that no specific assumption w.r.t. the character of $\xi$ is made in the context of this work.

\subsection{Truncation Error and Mappings}
The truncation error $
\rv{e}_\ell := \rv{\argu} - \proj{\argu}{\ell}
$
can be shown to be orthogonal to the \pce $\proj{\argu}{\ell}$, i.e.  $\rv{\argu} - \proj{\argu}{\ell} \perp \proj{\argu}{\ell}$. The error~$\rv{e}_\ell$ also satisfies
$
\underset{\ell \to \infty}{\lim}
\| \rv{e}_{\ell} \| = 0
$ \cite{Sullivan15book, Xiu10book, Cameron1947}.
Furthermore, if the weight to which the polynomials are orthogonal matches the (product) measure $\mu$, convergence of the above limit is known to be exponential \cite{Cameron1947, Wiener38, Xiu02}.
Besides exponential convergence several questions with respect to the truncation error are immediate:
First, is it possible to describe a random variable and its mapping precisely by a finite \pce?
Second, if the \pce is truncated early, is it possible to establish an error bound on $\rv{y}$?
To answer the above questions we define the minimum degree of a \pce as follows:

\begin{defi}[Minimum expansion degree]
	The minimum expansion degree of $\argu \in \Ltwospace$ is the number $d_{\arguargu} \in \mathbb{N}_0$ such that all \pce coefficients associated with higher-degree basis polynomials are zero, i.e. $\arguargu_j = 0$ for all $j$ with $\operatorname{deg} \phi_j > d_{\arguargu}$.
	\hfill $\square$
\end{defi}

\begin{ass}[Exact \pce input]
	\label{ass:ExactInput}
	For a given orthogonal polynomial basis $\{ \phi_j\}_{j=0}^{\ell_{\arguargu}}$, the \pce of the real-valued random variable $\argu \in \Ltwospace$ has the known and finite minimum degree $d_{\arguargu} \in \mathbb{N}_0$, and $\ell_{\arguargu} + 1$ \pce coefficients, cf. \eqref{eq:PCEDimensionFormula}.
	\hfill $\square$
\end{ass}
In other words, Assumption  \ref{ass:ExactInput} implies that a finite number of \pce coefficients yields a vanishing truncation error for the input uncertainty; i.e.,
\begin{equation}
\forall \ell \geq \ell_{\arguargu}: \quad \| \argu - \proj{\argu}{\ell} \| = 0.
\end{equation}
Indeed, for many applications Assumption~\ref{ass:ExactInput} is assumed to hold for a minimum degree of $d_z=1$; in other words, Gaussian, Beta, Gamma, or uniform distributions are employed to model uncertainties \cite{Mesbah14,Fagiano12a,Kim13a,Muehlpfordt17a,Mesbah16,Lucia15}.

We now turn back to the main question of quantifying the approximation error when describing the random variable $\rv{y}$ as a function of $\argu$ as given by \eqref{eq:mapping}. For the case of univariate Gaussian stochastic germ $\rv{\xi}$ the following result is known.
\begin{lemm}[Bound in univariate Hermite basis \cite{Augustin08}]
	\label{thm:Augustin}
	\mbox{}
	\noindent 
	Let $\nxi = 1$, and let the stochastic germ $\rv{\xi}$ be a standard Gaussian random variable.
	Consider $\argu$, $\rv{y}$ $\in \mathrm{H}$ $= \mathrm{L}^2({\mathbb{R},\mu_{\mathrm{Gauss}};\mathbb{R}})$, where $\rv{y} = f(\argu)$ with $f: \mathrm{H} \rightarrow \mathrm{H}$ square-integrable.
	Furthermore, let $f$ be $k$ times continuously differentiable with $f(\argu)^{(k)}\in \mathrm{H}$.
	Then, for $k \leq n+1$ it holds that
	\begin{equation}
	\label{eq:Error_Augustin}
	\left\| \rv{y} - \sum_{j=0}^{n} y_j \her{j} \right\| \leq \frac{\| f(\argu)^{(k)} \|}{ \prod_{i=0}^{k-1} \sqrt{n-i+1}} =: \eaug{k}{n},
	\end{equation}
	where $\her{j}$ is the $j$th probabilists' Hermite polynomial. \hfill $\square$
\end{lemm}
Lemma~\ref{thm:Augustin} provides  an error bound specifically tailored to univariate Gaussian uncertainties. 
However, obtaining sharp bounds for the general case is especially important for many applications in systems and control, specifically w.r.t. performance and stability results.
Two questions arise:
\begin{mylist}
	\item \label{item:EqualDimension} Choosing the \pce dimension equal to the \pce input dimension, i.e. $n +1 = \ell_{\arguargu} + 1$, what truncation error is made given a square-integrable nonlinear mapping $f(\cdot)$? 
	\item \label{item:MinimumDimension} What is the minimum \pce dimension $n + 1$ such that a zero truncation error is attained for $\rv{y}$?
\end{mylist}

\section{Truncation Errors for Polynomial Maps}
\label{sec:PolTruncationErrors}
Let  $\argu_i, \rv{y} \in \mathrm{H} = \Ltwospace$ with $i = 1, \hdots, n_{\arguargu}$ be real-valued random variables.
Moreover, let $\mathrm{\Argu} \subset \mathrm{H}$ be a complete subspace of dimension $n + 1$ generated by the orthogonal polynomial basis functions $\{\phi_j\}_{j=0}^{n}$.
For ease of presentation we assume that all $\argu_i$ have the same \pce dimension.
\begin{thm}[Error under polynomial mapping]	
	\label{thm:PolynomialMapping}
	Suppose that all $\argu_i$ satisfy Assumption \ref{ass:ExactInput} with minimum degree $d_{\arguargu}$ and a respective orthogonal basis, and let $f: \mathrm{H}^{n_{\arguargu}} \rightarrow \mathrm{H}$ be a square-integrable polynomial mapping of degree $d_f$ such that $\rv{y} = f(\argu_1, \hdots, \argu_{n_\arguargu})$.
	Then, the magnitude of the truncation error	$\rv{e}_n = \rv{y} - \proj{\rv{y}}{n}$ is
	\begin{equation}
	\label{eq:Error_PolynomialMapping}
	e_n := \| \rv{e}_n \| = 
	\begin{cases}
	\sqrt{\sum_{j = n + 1}^{\ell} y_j^2 \| \phi_j \|^2}, &  n < \ell, \\
	0, & n \geq \ell,
	\end{cases}
	\end{equation}
	where
	\begin{equation}
	\ell+1 = \frac{(\nxi + d_{\arguargu} d_f)!}{(\nxi! (d_{\arguargu} d_f)!)},
	\end{equation}
	and $y_j$ are the \pce coefficients of $\rv{y}$.\hfill $\square$
\end{thm}	

\emph{Proof: }
	The \pce for $\argu_i$ is given by
	\begin{equation}
	\argu_i = \sum_{j=0}^{\ell_{\arguargu}} \arguargu_{i,j} \phi_j,
	\end{equation}
	where $\ell_{\arguargu} + 1 = (\nxi + d_{\arguargu})!/(\nxi! d_{\arguargu}!)$.
Substituting this into the polynomial mapping $f(\cdot)$, one obtains
	\begin{equation}
	\label{eq:PolyMapping}
	\begin{aligned}
	\rv{y} &= f\Bigg(\sum_{j=0}^{\ell_{\arguargu}} \arguargu_{1,j} \phi_j,\, \hdots, \, \sum_{j=0}^{\ell_{\arguargu}} \arguargu_{n_{\arguargu},j} \phi_j\Bigg) \\
	&= \sum_{j=0}^{d_{\arguargu} d_f} \sum_{i=1}^{\nxi} \alpha_{ij} \xi_i^j = \sum_{j=0}^{\ell} y_j \phi_j.
	\end{aligned}
	\end{equation}
	The highest-degree polynomial term of $\rv{y}$ has degree $d_{\arguargu} d_f$, thus enlarging the basis by $\ell - \ell_{\arguargu}$ elements.
	The number of basis elements is given by \eqref{eq:PCEDimensionFormula} with $d \rightarrow d_{\arguargu} d_f$.
	The orthogonal projection of $\rv{y}$ onto $\mathrm{\Argu}$ yields $\proj{\rv{y}}{n} = \sum_{j=0}^{n} y_j \phi_j$.
	Consequently, the truncation error $\rv{e}_n$ becomes $\rv{e}_n = \sum_{j=n+1}^{\ell} y_j \phi_j$,
	which is zero in case of $n \geq \ell$.
	For $n < \ell$, apply Parseval's identity to obtain $\| \rv{e}_n \|$, cf. \cite{Bronstein13}.
	\QED \vspace*{2mm}

\noindent In light of Theorem \ref{thm:PolynomialMapping}, the answers to questions \ref{item:EqualDimension} and \ref{item:MinimumDimension}  are summarized.
\begin{coro}[Error for polynomial $f(\cdot)$]
\mbox{}
\begin{mylistlist}
	\item \label{item:Answer1} Given a polynomial mapping $f(\cdot)$ such that $\rv{y} = f(\argu)$ with $\argu,\rv{y} \in \Ltwospace$, and choosing the \pce dimension $\ell + 1$ equal to the \pce input dimension $\ell_{\arguargu} + 1$, the truncation error is given by $e_{\ell_{\arguargu}}$ from \eqref{eq:Error_PolynomialMapping}.\footnote{In order to evaluate \eqref{eq:Error_PolynomialMapping}, the terms $\| \phi_i \|^2$ have to be computed. This raises the question of the attached computational complexity. The terms $\| \phi_i \|^2$ can, for example, be computed using Gauss quadrature, which has a manageable computation complexity, e.g. $\mathcal{O}(n)$ with the approach from \cite{Townsend16}.}
	\item \label{item:Answer2} Furthermore, the minimum dimension is $\ell+1 = (\nxi + d_{\arguargu} d_f)!/(\nxi! (d_{\arguargu} d_f)!)$.
		\hfill $\square$
	\end{mylistlist}
\end{coro}
It is fair to ask for a comparison w.r.t. the error bound from Lemma~\ref{thm:Augustin}.
To provide insight into this question, consider the setting from Lemma~\ref{thm:Augustin} (univariate Gaussian stochastic germ) in combination with a polynomial mapping $f(\cdot)$.
\begin{table}
	\caption{Squared truncation errors $e_{d_{\arguargu}}$ and $\eaug{d_{\arguargu}+1}{d_{\arguargu}}$ for Example \ref{exmpl:QuadraticMapping}.}
	\vspace{-5mm}
	\label{tab:ProjectionErrorExample}
	\def\arraystretch{1.25}	
	\begin{center}
		\begin{tabular}{lllm{3.2cm}}
		\toprule
		$d_{\arguargu}$ & 1 & 2 & 3 \\
		\midrule
		$(e_{d_{\arguargu}})^2$ & $2  \arguargu_1^4$ & $24 \, \arguargu_2^2 (\arguargu_1^2 + \arguargu_2^2)$ & $480\, \arguargu_2^2\, \arguargu_3^2$ $+$ $24 \, (\arguargu_2^2 $ $+$ $9\, \arguargu_3^2$  $+ 2 \, \arguargu_1 \arguargu_3 )^2 + 720 \, \arguargu_3^4$  \\
		$(\eaug{d_{\arguargu}+1}{d_{\arguargu}})^2$ & $2  \arguargu_1^4$ & $24 \, \arguargu_2^2 (\arguargu_1^2 + 4 \arguargu_2^2)$ & $2400 \, \arguargu_2^2 \arguargu_3^2 + 24 \, (\arguargu_2^2 + 9 $  $ \arguargu_3^2 + 2 \, \arguargu_1 \arguargu_3)^2 + 10800 \, \arguargu_3^4$ \\
		\toprule
	\end{tabular}
	\vspace{\movefig}
\end{center}
\end{table}
\begin{exmpl}
	\label{exmpl:QuadraticMapping}
Let $n_{\arguargu} = 1$, $n_\xi = 1$, and let $\argu$ be a Gaussian random variable with mean $\mu$ and standard deviation $\sigma > 0$.
Consider the mapping $\rv{y} {=} \argu^2$.
If $\mathrm{\Argu} = \operatorname{span}  \{\her{0}, \her{1} \}$, i.e. the subspace $\mathrm{\Argu}$ is spanned by the first two Hermite polynomials, Assumption \ref{ass:ExactInput} is satisfied with $d_{\arguargu} = 1$. In other words, the \pce coefficients of $\argu$ are $\arguargu_0 {=} \mu$ and $\arguargu_1 {=} \sigma$.
Direct inspection shows that
$\rv{y} {=}  f(\argu) {=} (\mu^2 {+} \sigma^2)\her{0} {+} 2\sigma \mu \her{1} {+} \sigma^2 \her{2}$.
The error becomes $\rv{e}_1 {=} \sigma^2 \her{2}$ with norm $ e_1 {=} \sqrt{2} \, \sigma^2 $.
For derivatives $k {\in} \{1,2\}$ the respective error \eqref{eq:Error_Augustin} becomes $\eaug{1}{1} {=} \sqrt{2} \sigma \sqrt{\mu^2 +\sigma^2}$ ${\geq} \eaug{2}{1} {=} \sqrt{2} \sigma^2 {=} e_1$.
The minimum exact \pce degree for $\rv{y}$ is $d_{\arguargu} d_f {=} 2$.
Adding another basis function $\operatorname{span} \{\her{j} \}_{j=0}^2 {\supset} \mathrm{\Argu}$,
the projection error becomes zero.
Table~\ref{tab:ProjectionErrorExample} shows the squared norm of $\rv{e}_{d_{\arguargu}} {=} \rv{y} {-} \proj{\rv{y}}{d_{\arguargu}}$ for ascending input degree $d_{\arguargu}$ and symbolic \pce input coefficients $\arguargu_0, \hdots, \arguargu_{d_{\arguargu}}$.
Exactness of the error $\eaug{d_{\arguargu}+1}{d_{\arguargu}}$ can only be ensured in the case of $d_{\arguargu} {=} 1$.\footnote{
	It is worth asking for general conditions such that the error bound \eqref{eq:Error_Augustin} is tight. Specifically for $f(\argu) = \argu$, and $\argu = \her{n+1}$, the errors \eqref{eq:Error_PolynomialMapping} and \eqref{eq:Error_Augustin} with $k = 1$ are identical \cite[Corollary 2.2]{Augustin08}, namely $(n+1)!$.
General conditions are, however, beyond the scope of this paper.}
In the other cases shown, $\eaug{d_{\arguargu}+1}{d_{\arguargu}} > e_{d_{\arguargu}}$ holds.\hfill $\square$
\end{exmpl}

\subsection*{Uncertainty Quantification for Quadratic Programs}
\label{sec:UncQuadraticPrograms}

It is evident that Theorem~\ref{thm:PolynomialMapping} can be applied to discrete-time \lti systems subject to  uncertainties, whenever the state transition map is polynomial in the uncertainty.  
In the following, we focus on uncertain convex quadratic programs (\qps) due to their important role in systems and control.
For example, model predictive control for discrete-time \lti systems with convex polytopic constraints and a convex quadratic cost function is well-known to be equivalent to solving a \qp repeatedly online at each time instant \cite{maciejowski2002predictive,Rawlings2009MPC}.
Also, \qps are the basis for sequential quadratic programming methods for solving nonlinear programs that are encountered in nonlinear \mpc.
In many cases, however, the problem data of the \qp is uncertain---in these cases \pce is of advantage.

\begin{problem}[\qp with uncertain data]
	\label{prob:QPwithUncertainData}
	Let $\rv{h}$ be an $\mathbb{R}^{\Nx}$-valued random vector with elements $\rv{h}_1, \hdots , \rv{h}_{\Nx} \in \mathrm{H} = \Ltwospace$.
	Also, let $\rv{b}$ be an $\mathbb{R}^{\Ncon}$-valued random vector with elements $\rv{b}_1, \hdots, \rv{b}_{\Ncon} \in \mathrm{H}$. 
	Set $\argu = [\argu_1^\top ~ \argu_2^\top]^\top := [\rv{h}^\top ~ \rv{b}^\top]^\top$, where $\cdot^\top$ denotes the matrix transpose.
	All random variables satisfy Assumption \ref{ass:ExactInput}, each with known and exact finite \pce dimension $d$.\footnote{For the sake of brevity of notation, we demand the same dimension $d$.}
	Let $\mathrm{\Argu} \subset \mathrm{H}$ be a complete subspace of dimension $n + 1$ generated by the orthogonal polynomial basis functions $\{\phi_i\}_{i=0}^{n}$.
	Consider
	\begin{equation}
	\begin{aligned}
	\label{eq:QPProblemFormulation}
	& y := \argmin{\chi \in \mathbb{R}^{n_\chi}}   \frac{1}{2}\, \chi^\top \! H \chi + \arguargu_1^\top \chi \\
	& \phantom{\chi^\star = \,}\operatorname{s.t.} \quad ~~ A\chi + \arguargu_2 \leq 0,
	\end{aligned}
	\end{equation}
	for positive definite $H \in \mathbb{R}^{n \times n}$ and a non-empty feasible set $\{ \chi {\in} \mathbb{R}^{n_\chi} {:} A \chi {+} \arguargu_2 {\leq }0 \}$.
	The entries of the vectors $\arguargu_1 \in \mathbb{R}^{n_\chi}$ and $\arguargu_2 \in \mathbb{R}^{\Ncon}$ are realizations of the vector-valued random variables $\argu_{1}$ and $\argu_2$, respectively.
	Then, the problem is to find the $\mathbb{R}^{n_\chi}$-valued random variable $\rv{y}$ and quantify the element-wise truncation error $\| \rv{y}_i  - \proj{\rv{y}_i}{n}\|$ for all $i = 1, \hdots, n_\chi$.
	\hfill $\square$
\end{problem}

\begin{rema}[\qps and \mpc]
	In case of linear-quadratic \mpc, Problem \ref{prob:QPwithUncertainData} is equivalent to considering uncertainty with respect to the initial condition at every time instant \cite{maciejowski2002predictive}.
	This uncertainty  may be due to state estimation, or a lack of measurement precision/availability.
	\hfill $\square$
\end{rema}
\pce allows the influence of $\argu$ on $\rv{y}$ to be specified as follows.

\begin{thm}[Uncertainty quantification for convex \qps]
	\label{thm:QP_Uncertainty}
	For all realizations of $\argu$, let the active constraints in Problem~\ref{prob:QPwithUncertainData} satisfy the linear inequality constraint qualification (\licq) at the optimal solution $y$.
	\begin{enumerate}[label=(\roman*)]
		\item \label{item:ProofFirstItem}If the \pce dimension of $\rv{y}$ is chosen according to $n \geq d$, then its element-wise truncation error becomes zero, i.e. 
		\begin{equation}
		\| \rv{y}_i - \proj{\rv{y}}{n}_i \| = 0, \qquad i = 1, \hdots, n_\chi.
		\end{equation}
		\item \label{item:ProofSecondItem}If the \pce dimension of $\rv{y}$ is chosen as $n {<}  d$, and if the set of active constraints $\mathcal{A} = \{ a_1, \hdots, a_{\Nact} \} \subseteq \{ 1, \hdots, \Ncon \}$ is the same for all realizations of $\argu = [\rv{h}^\top ~ \rv{b}^\top]^\top$, then the element-wise truncation error becomes
		\begin{equation*}
		\| \rv{y}_i - \proj{\rv{y}}{n}_i \| {=} \sqrt{ \sum_{j = n + 1}^{d} \! \left( 
			w_i^{h \top} h_j +  w_i^{b \top}  \Mact b_j
			\right)^2   \! \| \phi_j \|^2 },
		\end{equation*}
		where $w_i^{h \top}$, $w_i^{b \top} $ are the $i$th rows with $i = 1, \hdots, n_\chi$ of the matrices $W^h$, $W^b$ that satisfy
		\begin{equation}
		\label{eq:InvertedCoefficientMatrix}
		\begin{bmatrix}
		W^h & W^b \\ V^h & V^b
		\end{bmatrix}
		= -
		\begin{bmatrix}
		H & A^\top \Mact^\top \\ \Mact A & 0
		\end{bmatrix}^{-1}.
		\end{equation}
		The active constraint selection matrix $\Mact \in \mathbb{N}_0^{\Nact \times \Ncon}$ is constructed from the active set $\mathcal{A}$ and has elements $(\Mact)_{ia_i} = 1$ for $i = 1, \hdots, \Nact$, zero elsewhere.
		\hfill $\square$
	\end{enumerate}	
\end{thm}

\emph{Proof: }
		Part (i)---Regardless of the realizations of $\argu$ there always exists a (possibly empty) set of active constraints $\mathcal{A}$ for the optimal solution $y = \chi^\star$. 
		Rewrite \eqref{eq:QPProblemFormulation} as
		\begin{equation}
		\begin{aligned}
		&\underset{ \chi \in \mathbb{R}^{n_\chi}}{\operatorname{min}} \quad \frac{1}{2}\, \chi^\top \! H \chi + \arguargu_1^\top \chi \\
		&\operatorname{s.t.} \quad \Mact A \chi + \Mact \arguargu_2  = 0,
		\end{aligned}
		\end{equation}
		where $\Mact$ selects the active constraints.
		The \kkt conditions become
		\begin{equation}
		\label{eq:KKTConditionsQP}
		\begin{bmatrix}
		H & A^\top \Mact^\top \\
		\Mact A & 0
		\end{bmatrix}
		\begin{bmatrix}
		y \\ \lambda^\star
		\end{bmatrix}
		=
		- \begin{bmatrix}
		\arguargu_1 \\ \Mact \arguargu_2
		\end{bmatrix}.
		\end{equation}
		Due to \licq the coefficient matrix is invertible, yielding exactly one solution.
		The argmin-operator maps the realizations $\arguargu$ linearly to $y$.
		Consequently, it maps the random variable $\argu$ linearly to the random variable $\rv{y}$.
		Thus, Theorem~\ref{thm:PolynomialMapping} is applicable with $d_f = 1$.
		
		Part (ii)---Because the set of active constraints is supposed to be $\mathcal{A}$ for all realizations, the \kkt conditions \eqref{eq:KKTConditionsQP} hold in terms of a function of random variables
		\begin{equation}
		\label{eq:argmin_operator}
		\begin{bmatrix}
		\rv{y} \\ \rv{\lambda}^\star
		\end{bmatrix}
		=
		\begin{bmatrix}
		W^h & W^b \\ V^h & V^b
		\end{bmatrix} \begin{bmatrix}
		\rv{h} \\ \Mact \rv{b}
		\end{bmatrix},
		\end{equation}
		where \eqref{eq:InvertedCoefficientMatrix} and $\argu = [\argu_1^\top ~ \argu_2^{\top}]^\top = [\rv{h}^\top ~ \rv{b}^\top]^\top$ are used.
		Invertibility follows again from \licq.
		Consequently for all $i = 1, \hdots, n_\chi$,
		\begin{equation*}
		\rv{y}_i = w_i^{h \top} \rv{h} +  w_i^{b \top} \Mact \rv{b} = \sum_{j=0}^{d} (w_i^{h \top} h_j +  w_i^{b \top} \Mact b_j) \phi_j,
		\end{equation*}
		and the result follows from Theorem \ref{thm:PolynomialMapping} with $d_f = 1$.
\QED \vspace*{2mm}
\begin{rema}[Extension to changes in the active set]
	Note that even if the active set changes, part i) of Theorem \ref{thm:QP_Uncertainty} still holds. Furthermore, the error description from part ii) can be turned into an upper bound by considering the
	\emph{worst case} active set, which maximizes $\| \rv{y}_i - \proj{\rv{y}}{n}_i \|$. Due to space limitations, we leave the details to future work. 
	\hfill $\square$
\end{rema}

Note that the computation of the \pce coefficients \eqref{eq:argmin_operator} is effectively \emph{instantaneous}, i.e. requires no significant computation time, whilst being \emph{exact}.
Arguably, this is not true for Monte Carlo or other sampling-based methods.
To illustrate Theorem \ref{thm:QP_Uncertainty} in use, we consider the following example.
\begin{exmpl}
	\label{exmpl:MPC}
	Consider linear-quadratic \mpc for an \lti discrete-time model $\chi(k+1) {=} A \chi(k) {+} B u(k)$ of an aircraft.
	The open-loop optimal control problem can be cast as a \qp \cite{maciejowski2002predictive}.
	The numerical values for the nominal system $(A,B)$ and weights $Q,R$ are taken from \cite{Lucia15}; the horizon length is $N = 35$.
	The input is the rate of change of the elevator angle, which introduces discrete-time integral action and an additional state.
	Uncertainty is introduced via the initial condition $\chi(0) = \argu$ for the altitude: it is modeled by the random variable $\argu_{4}$ that follows a Beta distribution on  $[-402, -381]$ with shape parameters $\alpha = 2$, $\beta = 5$, yielding the uncertain initial condition	$\argu = [0 ~  0 ~ 0 ~ \argu_{4} ~ 0]^\top$.
	 Assumption~\ref{ass:ExactInput} is satisfied with $d_{\arguargu} {=} \ell_{\arguargu} {=} 1$ and the \pce coefficients \timm{are} $\arguargu_{4,0} {=} - \!396$, $\arguargu_{4,1} {=} 3$ for a Jacobi polynomial basis.
	Following Theorem~\ref{thm:QP_Uncertainty}, a Jacobi polynomial basis with $n {\geq} 1$ allows a zero \pce truncation error in the decision variable $\rv{y}$.
	Figure \ref{fig:MPCexample} shows the evolution of the $6\sigma$-interval of the optimal input over time---note that the realization of the optimal random variable $\rv{y}(t)$ resembles the input $u$ to the system.
	For all realizations of the initial condition $\argu$ the constraints for the second state are active on the interval $[0.5, 7.5]$\,s.
	The corresponding optimal input trajectory over $[0.0, 7.0]$\,s is deterministic, as shown in Figure \ref{fig:MPCexample_6sigma}.
	In terms of \pce coefficients, this is equivalent to all \pce coefficients of order greater than zero being zero, yielding a Dirac-$\delta$-distribution.
	It is after the constraints become inactive that uncertainty plays a role; depicted in Figure \ref{fig:MPCexample_histogram} for $t {\in} [7.5, 17.0]$\,s.\footnote{The histogram rather than the \pdf is shown for sake of readability, because the peak of the \pdfs for $t \geq 12$ are orders of magnitude larger.}
	Because the closed-loop system is asymptotically stable, the input uncertainty eventually fades out, resulting again in Dirac-$\delta$-distributions.
\section{Truncation Errors for Non-Polynomial Maps}
\label{sec:NonPolTruncationErrors}
	\begin{figure}
	\centering
	\subfloat[Expected value and $6\sigma$-interval. Note that for $t \leq 7\,$s the input is deterministic as a state constraint is active.\label{fig:MPCexample_6sigma}]{\includegraphics{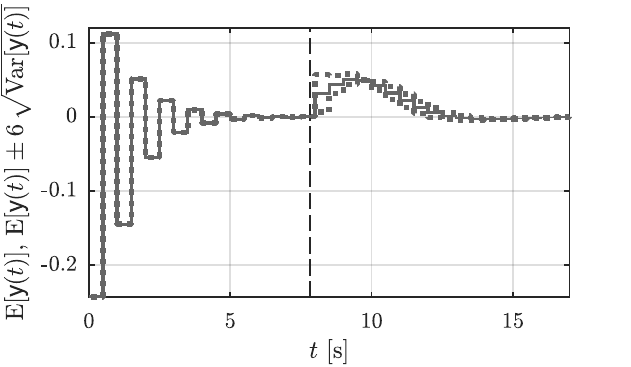}	}
	
	\vspace{-4.0mm}
	\subfloat[Histogram of optimal input for $t \geq 7.5\,s$.\label{fig:MPCexample_histogram}]{\includegraphics{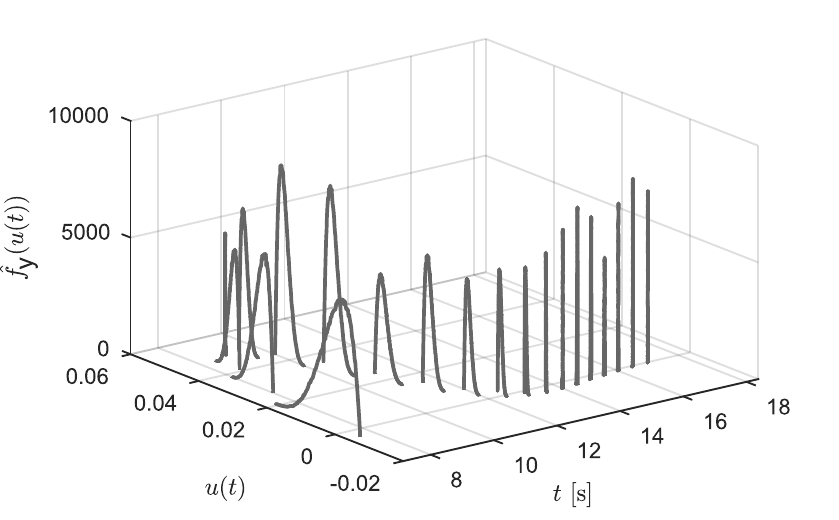}	
		\label{fig:ChanceConstraintLine394}}
	\caption{Optimal random variable input $\rv{y}$ for Example \ref{exmpl:MPC}.}
	\label{fig:MPCexample}
\end{figure}
We now turn to the question of  \pce truncation errors for non-polynomial mappings $f(\cdot)$.
Consider the real-valued random variables $\argu_i, \rv{y} \in \mathrm{H} = \Ltwospace$ with $i = 1, \hdots, n_{\arguargu}$.
Let $\mathrm{\Argu} \subset \mathrm{H}$ be a complete subspace of dimension $n + 1$ generated by the orthogonal polynomial basis functions $\{\phi_i\}_{i=0}^{n}$.
\begin{thm}[Error for non-polynomial mapping]
	\label{thm:NonpolynomialMapping}
	Let all $\argu_i$ satisfy Assumption \ref{ass:ExactInput} with minimum degree $d_{\arguargu}$ and a respective orthogonal basis, and let $f: \mathrm{H}^{n_{\arguargu}} \rightarrow \mathrm{H}$ be a square-integrable mapping such that $\rv{y} = f(\argu_1, \hdots, \argu_{n_{\arguargu}})$.
	Then, the magnitude of the truncation error	$\rv{e}_n = \rv{y} - \proj{\rv{y}}{n} $ is
	\begin{equation}
	\label{eq:LuenbergerError}
	e_n := \| \rv{e}_n \|  = \sqrt{ \| \rv{y} \|^2 - g^\top  Q g},
	\end{equation}
	where $Q = \operatorname{diag}(1/\| \phi_0 \|^2, \hdots, 1/\| \phi_n \|^2) \in \mathbb{R}^{(n+1) \times (n+1)}$ is positive definite, and $g = [g_1 ~ \hdots ~ g_{n+1}]^\top \in \mathbb{R}^{n + 1}$ with $g_{j+1} = \langle \rv{y}, \phi_j \rangle$ for all $j = 0, \hdots, n $.
	\hfill $\square$
\end{thm}

\emph{Proof:}
	The \pce coefficients of $\proj{\rv{y}}{n}$ satisfy
	\begin{equation}
	\label{eq:NormalEquations}
	\langle \phi_j, \phi_j \rangle y_j = \langle \rv{y}, \phi_j \rangle, ~ j = 0,\hdots,n ~ \Longleftrightarrow ~ Q^{-1} \bs{y} = g,
	\end{equation}
	which follows from orthogonality of the basis spanning $\mathrm{\Argu}$.
	The vector of \pce coefficients $\bs{y} \in \mathbb{R}^{n+1}$ contains all \pce coefficients $\bs{y} = [y_0 ~ \hdots ~ y_n]^\top$.
	The truncation error satisfies
	\begin{equation}
	\| \rv{e}_n \|^2 = \langle \rv{y}-\proj{\rv{y}}{n}, \rv{y}-\proj{\rv{y}}{n}\rangle = \| \rv{y} \|^2 - g^\top \bs{y},
	\end{equation}
	because $\rv{y} - \proj{\rv{y}}{n} \perp \proj{\rv{y}}{n} $.
	Using \eqref{eq:NormalEquations}, result \eqref{eq:LuenbergerError} follows. \QED \vspace*{1mm}

\noindent 
The error \eqref{eq:LuenbergerError} can be computed efficiently using Gauss quadrature.
In light of Theorem \ref{thm:NonpolynomialMapping}, the answers to questions \ref{item:EqualDimension} and \ref{item:MinimumDimension} are summarized.

\begin{coro}[Error for non-polynomial $f(\cdot)$]\vspace*{-.75mm}
	\mbox{}
\begin{mylistlist}[start=3]
	\item Given a non-polynomial mapping $f(\cdot)$ such that $\rv{y} = f(\argu)$ with $\argu,\rv{y} \in \Ltwospace$, and choosing the \pce dimension $\ell + 1$ equal to the \pce input dimension $\ell_{\arguargu} + 1$, the truncation error is given by $e_{\ell_{\arguargu}}$ from \eqref{eq:LuenbergerError}.
	\item No general statement is possible.\footnote{Note that in the univariate case $n_{\arguargu} = 1$ a zero truncation error in general requires an infinite \pce dimension, because a non-polynomial function cannot be represented exactly by a linear combination of a finite polynomial basis.} 
	However, for a user-specified error threshold the according minimum \pce dimension is obtained from Theorem \ref{thm:NonpolynomialMapping}.
		\hfill $\square$
\end{mylistlist}
\end{coro}
We illustrate Theorem \ref{thm:NonpolynomialMapping} for a continuous-time \lti example with \lqr and an uncertain system matrix.
\begin{exmpl}
	\label{exmpl:Aircraft}
	Consider the continuous-time \lti dynamics $\dot{\chi} = \rv{A}(\argu) \chi + B u$ for a modified aircraft model from \cite{maciejowski2002predictive}.
	The initial condition is $\chi(0) = [0 ~ 0 ~ 0 ~ 40]^\top$, and
	\begin{equation*}
	\rv{A} {=} 
	\begin{bmatrix}
	-1.2822{ +} 0.4\, \argu & 0 & 0.98 & 0 \\
	0 & 0 & 1 & 0 \\
	-5.4293 & 0 & -1.8366 & 0 \\
	-128.2 & 128.2 & 0 & 0
	\end{bmatrix}\! \!,
	B {=} \begin{bmatrix}
	-0.3 \\ 0 \\ -17 \\ 0
	\end{bmatrix}\!\!,
	\end{equation*}
	where $\argu \sim \mathrm{U}[-1,1]$.
	The realization $z=0$ corresponds to the nominal system matrix $A$.
	The control $u(t) = -K \chi(t)$ is computed via \lqr using the weights $Q = 0.001 \cdot \mathrm{I}_4$, $R = 100$ for the nominal system $(A,B)$.
	Now apply the above feedback to the uncertain system matrix.
	The closed-loop altitude trajectories $\chi_4(t)$ are given in Figure \ref{fig:ContinuousLTISystem} (left) for  best case and worst case realizations, clearly showing the performance degradation under uncertainty.
	The uncertainty~$\argu$ is mapped to the state $\chi(t)$ via the state transition map $\chi(t) = \operatorname{exp} [(\rv{A}(\argu)-BK)t] \, \chi_0$.
	Figure \ref{fig:ContinuousLTISystem} shows the altitude truncation error $e_{4,n}(t)$ from \eqref{eq:LuenbergerError} over time for increasing highest-degree $n \in \{2,3,4\}$.
	The basis consists of Legendre polynomials.
	The closed-loop system is asymptotically stable for all realizations of $\argu$, hence the truncation error decays to zero.
	However, it is clearly non-monotonic over time.
	Note how over- and undershooting of the deterministic solution, Figure \ref{fig:ContinuousLTISystem} left, carry over to the \pce error, Figure \ref{fig:ContinuousLTISystem} right.
	\hfill $\square$
\end{exmpl}

\begin{figure}
\centering
\includegraphics[]{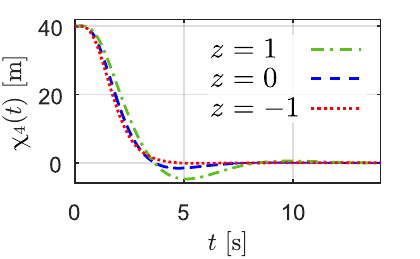}
\includegraphics[]{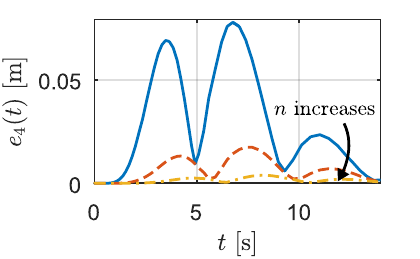}

\vspace{-3mm}
\caption{Closed-loop altitude trajectory for different $\argu$-realizations (left). Truncation error for altitude over time for difference \pce dimensions (right).}
\label{fig:ContinuousLTISystem}
\end{figure}

As to be expected, in case of a polynomial $f(\cdot)$, result \eqref{eq:Error_PolynomialMapping} is recovered and computationally cheaper.
\begin{lemm}
	Let $f: \mathrm{H}^{n_{\arguargu}} \rightarrow \mathrm{H}$ be a polynomial of degree $d_f$. Then \eqref{eq:LuenbergerError} is equivalent to \eqref{eq:Error_PolynomialMapping}.
	\hfill $\square$
\end{lemm}

\emph{Proof: }
	For polynomial $f(\cdot)$, the exact \pce for $\rv{y}$ is given from \eqref{eq:PolyMapping}.
	Consequently, for $\ell + 1 {=} (\nxi {+} d_{\arguargu} d_f)!/$ $(\nxi! (d_{\arguargu} d_f)!)$ we have
	\begin{align}
	\label{eq:ErrorInLemma}
	e_n^2 = \| \rv{y} \|^2 - g^\top Q g = \sum_{j=0}^{\ell} y_j^2 \| \phi_j \|^2 - \sum_{j=0}^{n} \frac{g_{j+1}^2}{\| \phi_j \|^2}.
	\end{align}
	For all $j {=}0, 1, \hdots, n$ with $n {\leq} \ell$, the numerators $g_{j+1}^2$ become
	\begin{equation}
	\label{eq:Numerators}
	g_{j+1}^2 = \langle \rv{y}, \phi_j \rangle^2 = \Big\langle  \sum_{i=0}^{\ell} y_i \phi_i, \phi_j \Big\rangle ^2 = 	(y_j \| \phi_j \|^2)^2.
	\end{equation}
	For $n > \ell$, the numerators $g_{j+1}$ are zero.
	Assuming $n \leq \ell$, use \eqref{eq:Numerators} in \eqref{eq:ErrorInLemma} to obtain
	$
	e_n^2 = \sum_{j=n+1}^{\ell} y_j^2 \| \phi_j \|^2,
$
	from which \eqref{eq:Error_PolynomialMapping} follows. \QED
\section{Conclusion \& Outlook}
Polynomial chaos  is an increasingly popular method for uncertainty propagation in systems and control.
Thus, the quantification of truncation errors stemming from the spectral expansions used in \pce is important.
For the case of polynomial and non-polynomial mappings---which might be state propagation maps of dynamic systems, algebraic equations or argmin-operators of convex problems---this work derived error bounds based on Hilbert space methods.
Specifically, the presented results provide an answer to the question of how to choose the \pce order such that the truncation error vanishes.
Several simulation results underpin the accuracy of the provided bounds and demonstrate how they can be used.
Future work will focus on non-convex optimization and optimal control problems.
\end{exmpl}

\printbibliography

\end{document}